\documentclass[journal]{IEEEtran}
\usepackage{tabularx}
\usepackage{cite}
\usepackage{color}
\usepackage{url}
\ifCLASSINFOpdf
\else
   \usepackage[dvips]{graphicx}
\fi
\usepackage[cmex10]{amsmath}
\usepackage[caption=false,font=footnotesize]{subfig}
\begin{document}
%
\title{A Circularly Symmetric Antenna Design With High Polarization Purity and Low Spillover}
%
%
%

\author{C.~M.~Holler,~A.~C.~Taylor,~M.~E.~Jones,~O.~G.~King,~S.~J.~C.~Muchovej,~M.~A.~Stevenson,\\~R.~J.~Wylde,~C.~J.~Copley,~R.~J.~Davis,~T.~J.~Pearson~and~A.~C.~S.~Readhead
\thanks{Manuscript received November 10, 2011; revised March 21, 2012}%
\thanks{This work was supported by NSF grants AST-0607857 and
  AST-1010024, and by the University of Oxford Department of Physics. A. C. Taylor acknowledges support from a Royal Society Dorothy Hodgkin Fellowship.}%
\thanks{C. M. Holler was with the Subdepartment of Astrophysics,
  Department of Physics, Oxford University, Denys Wilkinson Building,
  Oxford OX1 3RH, UK and is now with University of Applied
  Sciences Esslingen, Kanalstrasse 33, 73728 Esslingen am Neckar, Germany
  (phone: +49-7161-6791269 fax: +49-7161-6792177 e-mail:
  christian.holler@hs-esslingen.de).}%
\thanks{M. E. Jones and A. C. Taylor are with the Subdepartment of
  Astrophysics, Department of Physics, Oxford University, Denys
  Wilkinson Building, Oxford OX1 3RH, UK (e-mail:
  mike@astro.ox.ac.uk and act@astro.ox.ac.uk).%
}
\thanks{O. G. King, T. J. Pearson, A. C. S. Readhead and M. A. Stevenson are with the Department
  of Astronomy, California Institute of Technology, Pasadena, 91125,
  USA (e-mail: ogk@astro.caltech.edu, tjp@astro.caltech.edu,  acr@astro.caltech.edu and mas@astro.caltech.edu).%
}
\thanks{S. J. C. Muchovej is with the Owens Valley Radio Astronomy
  Observatory, California Institute of Technology, Pasadena, 91125,
  USA (e-mail: sjcm@astro.caltech.edu).}  \thanks{R. J. Wylde is with
  the School of Physics and Astronomy, University of St Andrews, Fife,
  UK and Thomas Keating Ltd, Billingshurst, West Sussex, RH14 9SH, UK
  (e-mail: r.wylde@terahertz.co.uk).
}
  \thanks{C. J. Copley was with the Department of Physics and Electronics, Rhodes University, Artillery Road, Grahamstown, South Africa and is now with the Subdepartment of Astrophysics, Department of Physics, Oxford University, Denys Wilkinson Building, Oxford OX1 3RH, UK (e-mail: Charles.Copley@astro.ox.ac.uk).
}
  \thanks{R. J. Davis is with the Jodrell Bank Centre for Astrophysics, University of Manchester, Manchester, M13 9PL, UK (e-mail: rjd@jb.man.ac.uk).
}}%

\markboth{\copyright~2012 IEEE Transactions on antennas and propagation}%
{Shell \MakeLowercase{\textit{et al.}}}
%

\maketitle

\begin{abstract}
\boldmath We describe the development of two circularly symmetric
antennas with high polarization purity and low spill-over. Both were
designed to be used in an all-sky polarization and intensity survey at
5 GHz (the C-Band All-Sky Survey, C-BASS). The survey requirements
call for very low cross-polar signal levels and
far-out sidelobes. Two different existing antennas, with 6.1-m and
7.6-m diameter primaries, were adapted by replacing the feed and
secondary optics, resulting in identical beam performances of
0.73$^\circ$~FWHM, cross-polarization better than $-50 \, \rm dB$, and
far-out sidelobes below $-70\, \rm dB$.  The polarization purity was
realized by using a symmetric low-loss dielectric foam support
structure for the secondary mirror, avoiding the need for secondary
support struts. Ground spill-over was largely reduced by using
absorbing baffles around the primary and secondary mirrors, and by the
use of a low-sidelobe profiled corrugated feedhorn. The 6.1-m antenna
and receiver have been completed and tested. Results
  show that the co-polar beam matches the design simulations very
  closely in the main beam and down to levels of $-80 \, \rm dB$ in
  the backlobes. With the absorbing baffles in place the far-out ($>
  100^{\circ}$) sidelobe response is reduced below $-90 \, \rm
  dB$. Cross-polar response could only be measured down to a noise
  floor of $-20 \, \rm dB$ but is also consistent with the design
  simulations. Temperature loading and groundspill due to the
  secondary support were measured at less than $1\, \rm K$.
\end{abstract}


%
\IEEEpeerreviewmaketitle

\section{Introduction}
%
%
%
%
\IEEEPARstart{T}{otal-power} radio-astronomical imaging, in which an
image is formed by scanning a single antenna beam over the sky,
presents some of the most stringent possible requirements on antenna
design. Accurate measurements require extreme stability of the
receiver system as well as minimization of scan-synchronous pickup via sidelobes and cross-polarization response. Such effects are
difficult or impossible to remove in post-processing, and drive the
need to limit the intrinsic sidelobe response and cross-polarization
of the antenna used for total-power imaging. Interferometers are less
sensitive to these effects and are typically used for high-resolution
radio imaging, but they are insensitive to scales larger than the
antenna beam. Large-scale (and in particular, all-sky) images are only
attainable by scanning single-dish telescopes. In this paper, we
describe the development of two antennas for a sensitive, all-sky
imaging experiment, which are designed to have very low levels of
sidelobes and cross-polarization.

The C-Band All-Sky Survey (C-BASS) \cite{cbass_king}, is
a project to image the whole sky at 5~GHz in both intensity and
polarization, in order to meet a variety of scientific goals in the
study of the cosmic microwave background (CMB) radiation and emission
processes in the Galaxy. C-BASS will be the first survey of diffuse
Galactic emission at a frequency low enough to be dominated by
synchrotron radiation but high enough to be largely uncorrupted by
Faraday rotation effects. The maps produced by the survey will enable
accurate subtraction of foreground contaminating signals from
higher-frequency CMB polarization sky surveys, including the
\emph{WMAP} \cite{WMAP} and \emph{Planck} \cite{Planck} satellites,
and will also be a major resource for studying the interstellar medium
and magnetic field of the Galaxy.

The C-BASS receiver is a combination of a broad-band (4.5--5.5 GHz) correlation polarimeter and a pseudo-correlation radiometer (also sometimes known as a continuous-comparison radiometer). The
polarimeter correlates right (RCP) and left (LCP) circular
polarizations to measure the Stokes $Q$ and $U$ parameters, while the
radiometer differences RCP and LCP separately against
temperature-stabilized loads to allow accurate total power (Stokes
$I$) measurements.

To obtain full-sky coverage, similar receivers are mounted on two
separate telescopes: one at the Owens Valley Radio Astronomy Observatory in
California, and the other at Klerefontein, near the MeerKAT Observatory, in South Africa. The survey resolution is 0.73$^\circ$, with a design
sensitivity of $< 0.1$ mK map rms in Stokes $I$, $Q$, and $U$. Since the
contrast between sky and ground temperatures is several hundred
Kelvins, limiting scan-synchronous sidelobe pickup to below the thermal
noise level requires that the antennas provide at least 60 dB of
rejection in the horizon direction. In addition, given that the
polarization fraction of the sky signal may be less than one percent,
obtaining accurate polarization angles requires the cross polarization
level (after calibration) to be better than $-$30~dB. 

\section{Antenna overview and optics design} 

Due to the availability of suitable antennas at each site, C-BASS uses
different antenna designs for the northern and southern surveys. The
northern antenna, which was originally designed as a prototype for the
NASA Deep Space Network \cite{Imbriale}, has a 6.1-m primary reflector
and was donated to the project by the Jet Propulsion Laboratory. The
southern antenna was designed by Mitsubishi for a low-earth-orbit
telecommunications satellite ground station. It has a 7.6-m diameter
primary mirror, and was donated by Telkom SA Ltd. Both antennas are
circularly symmetric and have shaped profiles. The 6.1-m antenna was originally optimized for maximum forward gain using
  uniform aperture illumination of the primary mirror. The original
  shaping criteria for the other antenna are unknown.  Photogrammetry
was used to measure the primary surface shape for both antennas. Since
both antennas were originally designed for satellite communication
purposes at different frequencies and were not
suitable as imaging instruments, the feed and secondary optics were
redesigned for this project.

We used Galindo's theory of shaped dual reflectors \cite{Galindo_1964}
for the redesign of the secondary mirrors. This provides a set of
three differential equations which are solved for the desired boundary
conditions. As no closed-form solution is possible, a simple program
was written to solve these equations iteratively. 
The optical design requirements for this astronomical survey were
 common main beam patterns (up to the first null) from both antennas, low sidelobe and cross polar levels, low blockage and scattering due to the secondary
mirror, constant phase in the aperture of the primary mirrors, and, if possible, a common
feed horn design for both antennas. 

\begin{figure}
\includegraphics[width=4.8in]{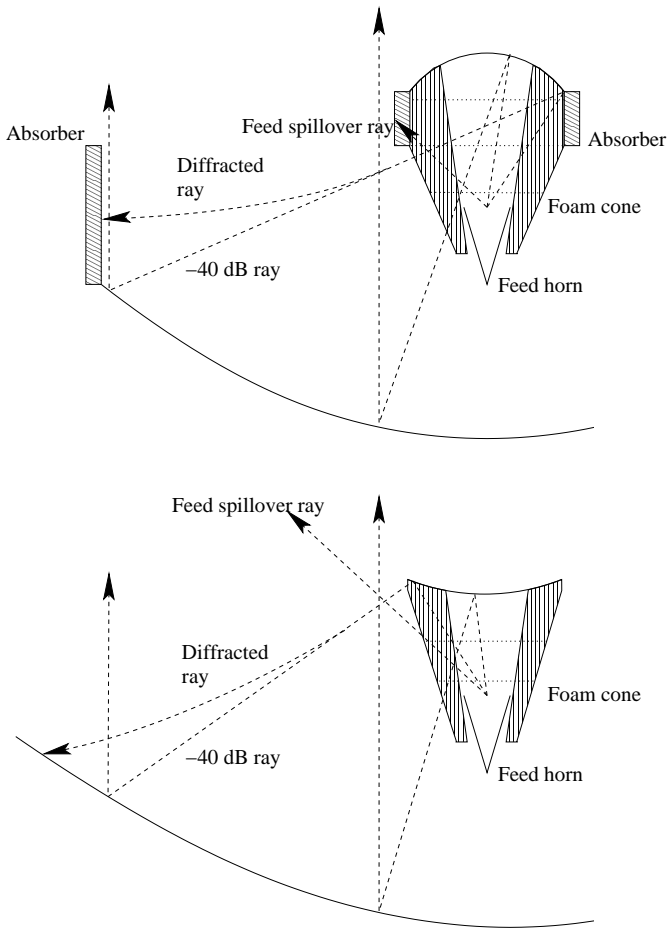}
\caption{Schematic ray diagrams (not to scale) of (top) the northern
  6.1-m antenna and (bottom) the southern 7.6-m antenna. In the
  northern antenna both direct spillover from the feed and diffraction
  spillover around the primary mirror edge are blocked by absorbing
  baffles. In the southern antenna, primary spillover is limited by the
  under-illumination of the larger antenna; direct spillover past the
  secondary cannot be further limited due to the Cassegrain
  geometry. Dotted lines in the foam cones indicate glue joints which
  are traversed by free-space rays. }
\label{rays}
\end{figure}

 The design procedure started with a ray optics
   analysis in order to obtain approximate secondary and feed aperture
   sizes. The first criterion was to limit direct spillover from the
   feed to the sky and the shadowing of the optics by the feed horn
   itself, which is to be less than the blockage shadow due to the
   secondary mirror. The different shapes of the primary surfaces,
   taken together with these requirements, necessitated different
   optical configurations -- Gregorian (concave ellipsoidal secondary)
   for the northern antenna and Cassegrain (convex hyperboloidal
   secondary) for the southern antenna. It was possible, however, to
   use a common feed horn design.  Fig. \ref{rays} shows a schematic
   ray diagram of each design.  In both cases our requirements
   together with the wavelength of the survey caused the diameter of
   the secondary to be significantly increased compared to the
   original optics -- the new mirrors each subtends a semi-angle of
 46.5$^\circ$ from the feed phase centre, at which angle the feed
 illumination taper is $-$40~dB at the band centre of 5~GHz.

 In the second step the design was optimized using
   Galindo's design equations. A manual iterative process was used to
   produce very similar main beam patterns for the two telescopes
   together with very low sidelobes across the full beam.  The cross
   polar level was relatively independent on the exact design and was
   mainly set by the cross polar signal level of the feed horn and the
   fact that our design has unbroken circular symmetry.

\begin{figure}[t]
\center
\includegraphics[width=3.5in]{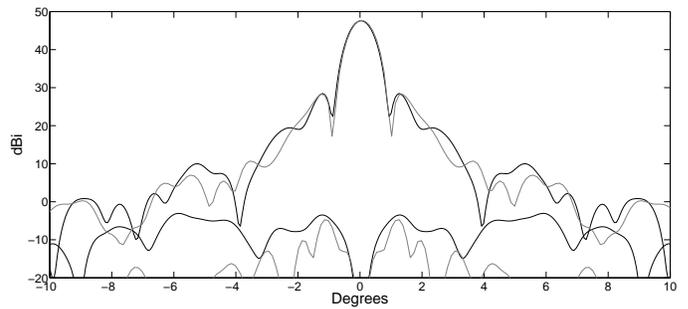}
\caption{Simulation of the co- and cross-polar forward beams of the
  6.1-m antenna (black) and the 7.6-m antenna (grey).}
\label{fig:near-beams}
\end{figure}

Fig. \ref{fig:near-beams} shows the simulated near-in
co-polar and cross-polar beams of the two antennas. The beams of the
two systems are well matched, allowing for the northern and southern
surveys to be combined together with no change of resolution. The main
beam efficiency (i.e., the fraction of the total radiated power within
the first null) is calculated to be 80\%, while the power within
$5^{\circ}$ (which encompasses most of the diffraction due to the
secondary blockage) is 91\%. This compares favourably with antennas
used for previous total-power surveys, e.g. the Stockert 25-m antenna
\cite{reich82} which has a main-beam efficiency of 55\% and a
`full-beam' (within $7^{\circ}$) efficiency of 69\%. The high beam
efficiency of the C-BASS antennas is due to the lack of scattering
structures within the beam, and will result in a much simpler and more
accurate conversion from measured power data to true sky temperatures. 

\section{Antenna design features}

To minimize the cross-polar response and far-out sidelobes, three
features have been introduced in to the antenna design:
\renewcommand{\labelitemi}{$-$}
\begin{itemize}
\item{A very low-sidelobe corrugated feed horn has been designed.}
\item{The secondary mirror is supported by a circular symmetric dielectric foam structure instead of support struts.}
\item{In the case of the 6.1-m antenna, absorbing baffles are mounted around the primary mirrors and also around the secondary mirror. In the case of the 7.6-m antenna, a very high degree of under-illumination is used.}
\end{itemize}

\subsection{Feed horn}
The feed horn used for both antennas is a custom design manufactured
by Thomas Keating Ltd., with highly suppressed sidelobe levels and low
cross-polarization. Drawing upon ideas designed to minimize
reflections in quasi-optical circuits \cite{Cruckshank_2007}, the
corrugated horn has a cosine-squared profile designed to generate a
significant amount of HE12 mode which is followed by a phasing section
which brings the HE12 mode into phase with the dominant HE11 mode. The
combination of these modes provides an aperture field with low edge
taper which translates into a pattern with very low sidelobes. 

\begin{figure}[t]
\center
\includegraphics[width=3.5in]{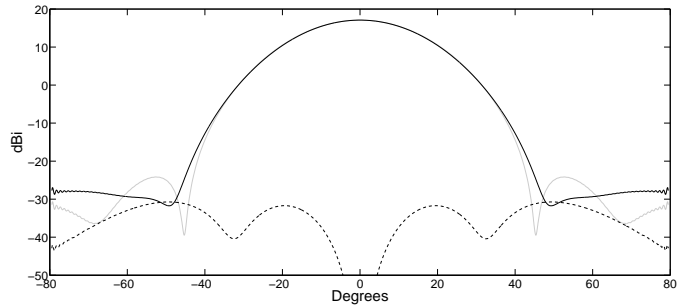}
\caption{Simulated feed horn pattern at 5 GHz; E-plane (grey), H-plane (black) and cross-polar (dotted black).}
\label{fig:horn-pattern}
\end{figure}

Modal matching predictions of the low sidelobe horn
  for C-BASS from CORRUG \cite{corrug} as well as modelling and
  measuring patterns for other similar horns at much higher
  frequencies \cite{Keating} show that this design of corrugated horn
  retains its very low sidelobe and cross-polar performance over large
  bandwidths. Our design is optimized for 5.0 GHz, at which the peak
  sidelobe level is $-40$~dB relative to the boresight level and the
  peak cross-polarization is $-47$~dB.  At the band edges at 4.5 GHz
  and 5.5 GHz the sidelobe levels degrade to $-32$~dB and
  cross-polarization to $-33$~dB, which is still acceptable
  performance over a 20\% bandwidth. The pattern of the feed horn at
  5.0 GHz, as predicted by CORRUG, is shown in
  Fig. \ref{fig:horn-pattern}.

\begin{figure}
\center
\includegraphics[trim=2.3cm 0cm 2.3cm 0.8cm, clip=true,width=3.5in]{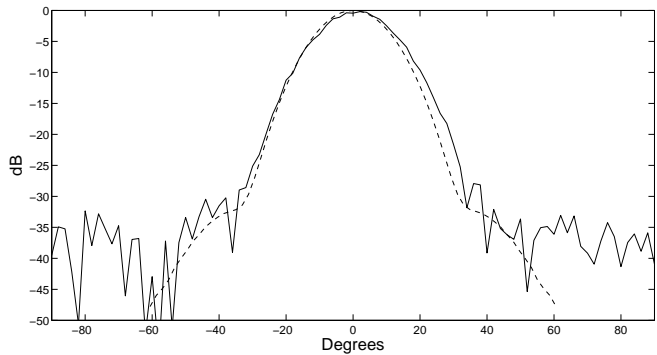}
\caption{Feed horn near-field (3~m) pattern at 4.5 GHz; measurement
  (black), simulation (dotted black).}
\label{fig:horn-measurement}
\end{figure}

The horn aperture radius is 113 mm,
and it produces an illumination at the edge of the secondary mirror of
$-40$ dB.  The horn is manufactured in three corrugated segments,
which are bolted to the outside of the receiver cryostat, and two
smooth conical segments which are integral to the outer (room
temperature) and intermediate (40~K) shells of the cryostat. The outer
smooth section also incorporates a Mylar vacuum window and a
Zotefoam plug to block infrared loading. The horn terminates in a
specially designed four-probe orthomode transducer \cite{Grimes_2007}
which generates left and right circular polarizations. In our facilities we could only
measure the near-field beam pattern at 3 m distance. This measurement is shown alongside the equivalent near-field simulation in
Fig.~\ref{fig:horn-measurement}.

\subsection{Secondary mirror and support}

In a circularly symmetric design, secondary support struts block and scatter signals in a manner which is difficult to control. The break in symmetry due to the struts results in elevated levels of cross-polar signals. In addition, the
scattered signals increase the system noise via ground pickup in a way
which is dependent on the pointing direction of the antenna. For
sensitive polarization experiments it is essential to minimize the
effect of the secondary support. One option is to use offset antennas
with no blocking. However, this can increase costs significantly,
especially since in our case two circularly symmetric antennas were
readily available. An alternative method is to support the mirror
using a material which is effectively transparent at the frequency of
operation, i.e., has both low loss and low refractive index. We chose
the material Plastazote from Zotefoams Ltd, which at low microwave
frequencies is almost fully transparent and therefore can be used to
support the secondary mirror without blocking or scattering the
signal.

\begin{figure}[!t]
\centerline{\includegraphics[scale=0.25]{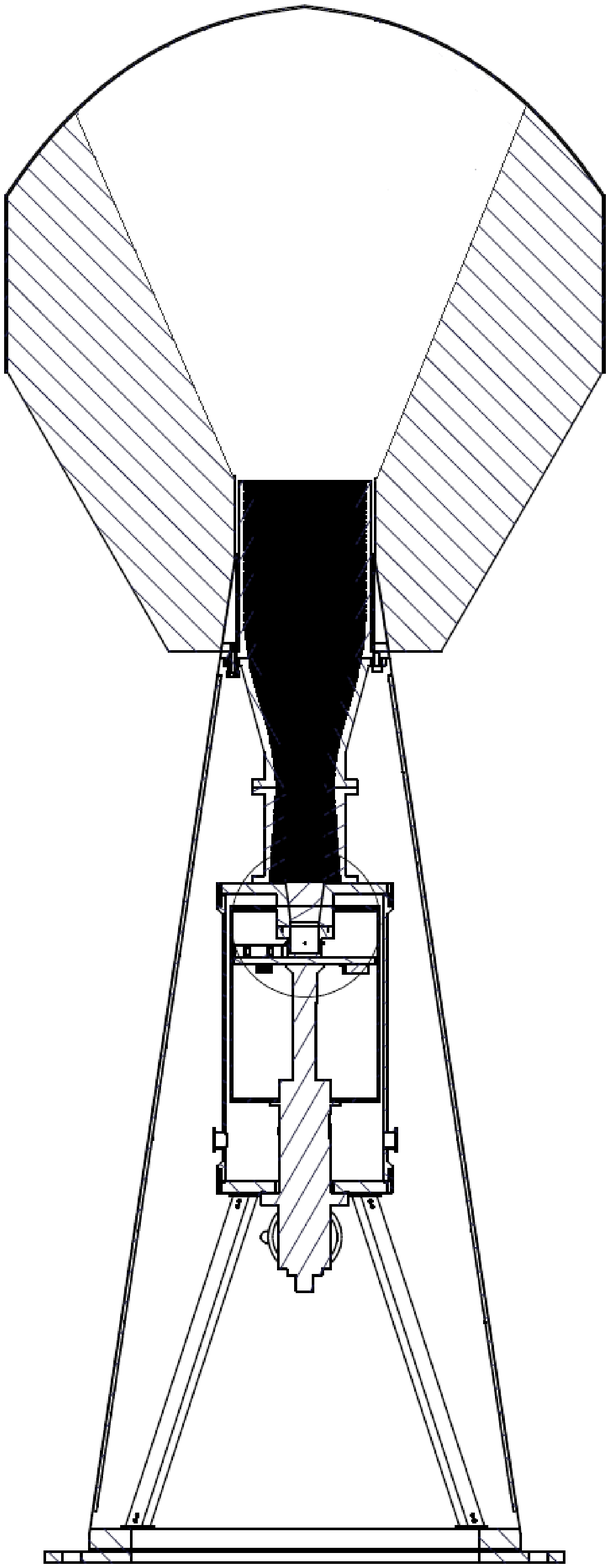}%
\hfil
\includegraphics[scale=0.48]{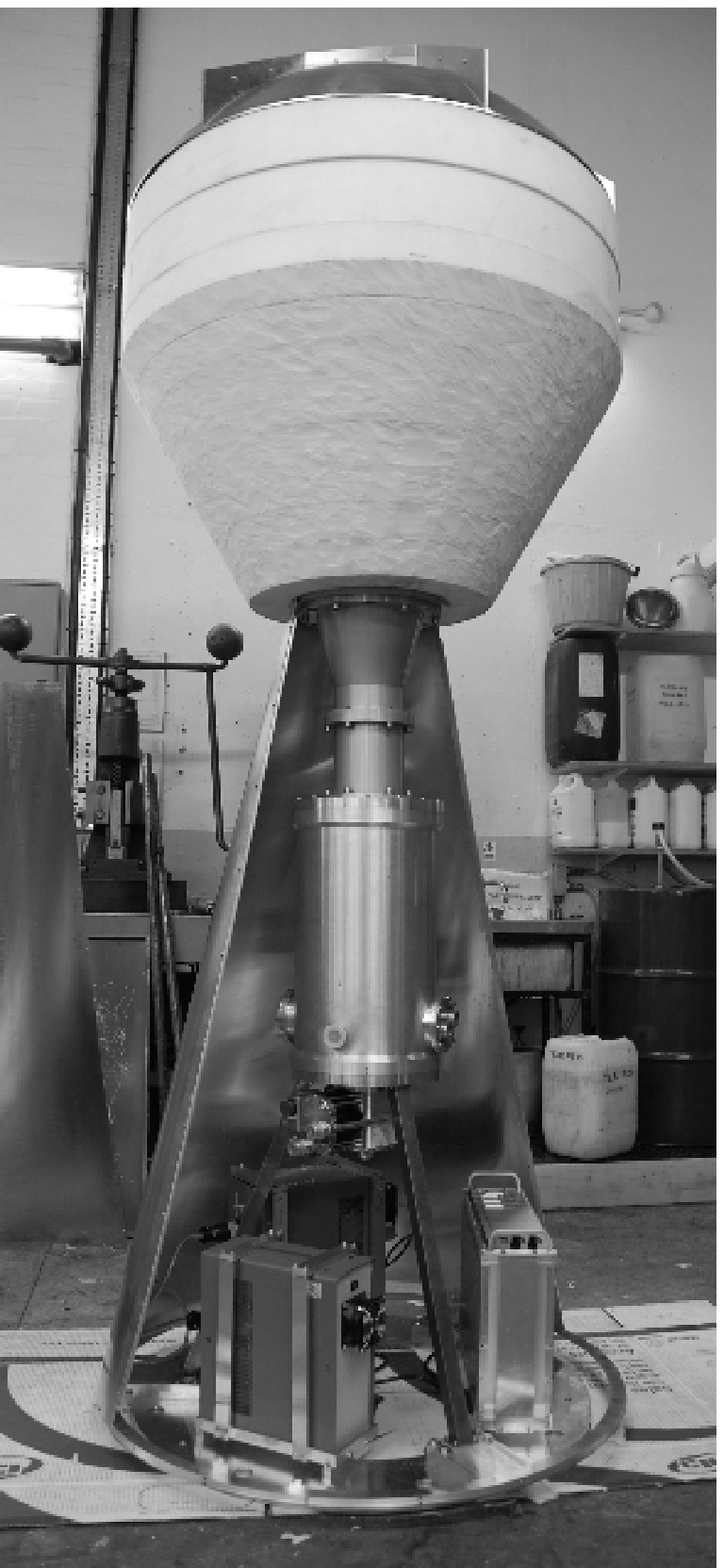}}%
\caption{Receiver, feed horn, secondary support cone and secondary mirror in a drawing (left) and during assembly (right).}
\label{fig:cone}
\end{figure}


\subsubsection{Secondary support material characteristics}

Plastazote is a closed-cell expanded polyethylene foam formed by the
vacuum expansion of heated polyethylene sheets that have had nitrogen
diffused into the solid phase under pressure. Two different grades are
available, using either low density (LD) or high density (HD)
polyethylene as the starting material. HD polyethylene (HDPE) is only
slightly denser (930 -- 970 kg m$^{-3}$ than LDPE (910 -- 930 kg m$^{-3}$)
but is considerably stiffer due to reduced branching of the polymer
chains and consequently higher intermolecular forces. Depending on the
amount of infused nitrogen, the foam density after expansion can be
between 15 and 115 kg m$^{-3}$.  The dielectric constant of solid LDPE
and HDPE is about 2.3 and the loss tangent, $\tan \, \delta$, is
$3.1\times10^{-4}$. We measured the dielectric constant and loss of
samples of the foam for various density and material grades by filling a
rectangular waveguide section with a foam sample and measuring the
change in phase and amplitude compared to an empty waveguide, using a
vector network analyzer. We found that the results were consistent
with a simple scaling of the properties with solid fraction of the
foam, i.e., the dielectric constant was given by
$$
\varepsilon_{\rm foam} -1 = (\rho_{\rm foam}/\rho_{\rm solid})(\varepsilon_{\rm solid} -1) \\
$$
and the loss tangent was given by 
$$
\rm{\tan}(\delta)_{\rm foam} = (\rho_{\rm foam}/\rho_{\rm
  solid})\rm{\tan}(\delta)_{\rm solid}.
$$ 

The foam can be glued to itself and to metals with standard adhesives,
and can also be heat welded to itself. Heat welding is used to build
up the basic manufactured sheets, which are around 30 mm thick, into
laminated blocks of thickness up to 300 mm (limited by the size of the
heat welding equipment). We used an impact adhesive (Evostick TX528)
to join together blocks to make larger structures. This adhesive was
tested for RF properties by inserting glued foam samples into a
waveguide, with the glue joint across the waveguide, and testing on a
vector network analyzer. No change in average transmission compared to
a single unglued piece of foam was detected at the level $< $0.05~dB.

\subsubsection{Secondary support design}

A drawing of the receiver, secondary support cone and secondary mirror
for the 6.1-m Gregorian system is shown in
Fig. \ref{fig:cone}. Initially, the foam cone was constructed out of
120-mm thick sheets of LD45 (LDPE base material with a density of 45
kg~m$^{-3}$), glued together in layers perpendicular to the optic
axis, and cut to shape with a bandsaw. This grade of foam was used for
the secondary mirror supports in the Cosmic Background Imager~2
instrument\cite{CBI2_2011}. The foam cone was glued (using impact
adhesive) to a hollow aluminium conical insert which mounts to a
flange on the outside of the feed horn. Adjustment screws at this
interface allow a small amount of motion in focus and tip-tilt. The
secondary mirror was glued to the top of the foam cone. 

After approximately 18 months in service it became apparent that some
of the foam-to-foam and foam-to-metal joints were beginning to fail
due to weathering and mechanical load. We decided to re-make the cone
using HD grade foams to increase the stiffness of the structure, and
to reduce the load on the most vulnerable joints. The base section of
the new cone, which bears the largest mechanical forces and transmits
them to the telescope structure, was made from a single piece of
HD115, which was available in heat-laminated blocks 300 mm
thick. HD115 has a flexural modulus of 23 MPa, compared to 1.0 MPa for
LD45. The remainder of the cone was made from blocks of HD30, which
has a flexural modulus of 3.6 MPa, higher than that of LD45 despite its
lower density. The base section was glued to the aluminium insert
using Araldite 2005 space-filling epoxy resin, which is stronger and
more weather-resistant than the impact adhesive. Epoxy was also used
for fixing a new carbon-fibre secondary mirror (see below) to the top
of the cone. The epoxy was measured to have significantly higher RF
loss than the impact adhesive; this is not important for the base
joint (which is outside the beam) or the mirror joint (which is a
reflective surface and therefore has zero electric field close to the
surface), but would be significant for the foam-to-foam joints which
the beam crosses in free space. We therefore retained the low-loss
impact adhesive for these joints but protected the joint edges against
weathering by wrapping the exposed cone structure in a thin
polyethylene membrane. Due to the thicker foam sections being
available only three such joints were needed, compared to eight in
the previous design. Using lower density foam for the bulk of the
structure and a lighter mirror also significantly reduced the stress
on the joints.

\subsubsection{Secondary mirror design}

Secondary mirrors of different shapes were required for the two antennas -- a
deep concave ellipsoidal mirror for the 6.1-m antenna, and a shallow
convex hyperboloidal mirror for the 7.6-m antenna. To minimize the
weight that has to be supported by the foam structure, a lightweight
design was developed. Initially for the 6.1-m antenna a
light-weighted machined aluminium mirror was used. The mirror was
constructed from seven aluminium segments, a central section and six
identical petals, bolted together via backing ribs. The back and sides of
each mirror segment were machined down to 3 mm thickness, and the
whole mirror had a mass of 12 kg. This mirror was used with the
initial foam support made from LD45. When this support was re-made
using HD foam we also decided to change to a carbon-fibre mirror, in
order to further reduce the mechanical load. The mirror was made to
exactly the same design as the first mirror, using five layers of
prepreg carbon fibre weave for each segment, laid up on a machined
aluminium mould, and cured under pressure in an
autoclave. Reflectivity measurements on carbon fibre samples showed a
reflection loss 1 per cent worse than solid aluminium, which would
have added about 3 K to the total system temperature, a significant
effect. We therefore covered the front surface with 3M aluminium
adhesive tape to provide a metallic reflecting surface. The total mass
of the carbon fibre mirror was about 2 kg. The maximum deflection of this new cone/mirror assembly under gravitational load was less than 0.1~mm, which corresponds to a shift in beam pointing of 10~arcsec. This is negligible compared to the beam size of 0.73$^{\circ}$. For the 7.6-m antenna secondary we follow the original manufacturing method of directly machining seven mirror segments from aluminium. This is significantly cheaper for a one-off production and, in combination with the HD foam, results in negligible shift in the beam pointing due to gravitational loading.

\subsubsection{Performance}

Possible contributions to the overall system noise from the foam cone
were estimated from measurements using the C-BASS receiver. The total
system noise power out of the cryogenic receiver was measured with the
bare feedhorn pointing at the sky, with the secondary optics in place,
and with the feed terminated with a room-temperature
absorber. The noise power contributions of the
  receiver, sky and optics were estimated from comparisons between the
  power measured from the sky, a room-temperature absorber, and
  absorber cooled to $77 \, \rm K$ with liquid nitrogen.  The excess
power attributable to the secondary optics (including loss in the foam
and ground pick-up from spillover around the primary) was no more than
$1\, \rm K$. Limits on the scattering produced by the foam are given
by the agreement between measured and calculated far-out sidelobes
(see Section IV below). The measurements were repeated for the new
foam cone made from HD foam, with identical results. If we had used a 
conventional secondary support system, GRASP-9
  simulations of a representative 4-strut support show that the
  expected blockage would be around $5 \%$ of the total power,
  resulting in a spillover contribution of 5 -- 7 K, which would be
  highly position-dependant and would result in significant
  scan-synchronous pickup during survey observations.

\begin{figure}[!t]
\center
\includegraphics[width=2.8in]{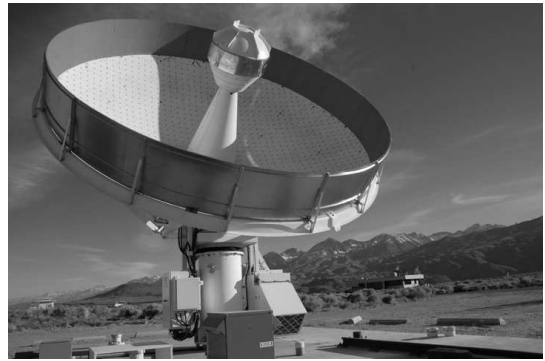}
\caption{The 6.1-m antenna at the Owens Valley Radio Astronomy
  Observatory with the new secondary support structure and the
  absorber-lined baffle around the rim of the primary and around the
  upper part of the foam secondary support structure.}
\label{fig:antenna}
\end{figure}

\subsection{Absorbing baffles}

\begin{figure}[t]
\center
\includegraphics[width=3.5in]{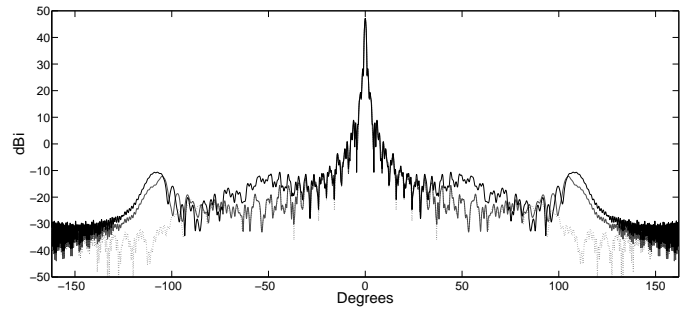}
\caption{Simulated beam pattern of the 6.1-m antenna at 4.5 GHz, without baffle
  (black), with secondary baffle only (grey) and with both baffles
  (dotted black) attached.}
\label{fig:baffles}
\end{figure}

Signal spillover past the primary and secondary mirrors produces
far-out sidelobes. This can increase the system temperature when these
sidelobes point at bright sources or the ground. In addition, these
sidelobes can be strongly polarized and induce changes in the intensity and
polarization signal as the pointing of the antenna
varies. Far-out sidelobes due to spillover past the edge of the
primary mirror can be significantly reduced by mounting an
absorber-lined cylindrical baffle around the rim of the primary mirror
\cite{Dybdal_1979}. We have taken this idea further and also fitted
the secondary mirror (on the 6.1-m antenna) with an absorbing baffle
to block spillover past the secondary mirror direct to the sky.  The
structure is shown in Fig. \ref{rays} and Fig. \ref{fig:antenna}. Due
to the different ray geometry a baffle around the secondary is only
possible with a Gregorian system and not with a Cassegrain system.

The absorbing material radiates at ambient temperature and therefore
increases the system temperature. However, this temperature component
is stable with pointing and unpolarized, and therefore produces no
changing signal as the antenna scans the sky. Provided the fraction of
the total illumination intercepted by the absorbers is kept small, the
benefit of reducing scan-synchronous pickup outweighs the increase in
system noise.

We found that a reduction in spillover sidelobe level  of 17~dB at 4.5~GHz
could be achieved with an absorber of height 0.8~m above the rim of
the 6.1-m primary. This results in a temperature load contribution of
0.7~K. The temperature contribution from the secondary baffle is
0.2~K. For the 7.6-m antenna this would be considerably less, 0.1~K
from the primary baffle, due to the lower edge illumination of the
larger antenna. In practice the benefit of the absorbing baffle for
the 7.6-m antenna is marginal given the very strong under-illumination
of the primary, and we have elected not to implement it.

The full beam patterns for the two antennas including the effect of
the absorbing baffles on the 6.1-m antenna were simulated using
GRASP9. To allow reasonable meshing of the surfaces representing the
absorbers in GRASP9 it was necessary to slightly displace them from
the mirror surfaces. This is not expected to have any significant
effect on the calculated beams. The characteristics of both antennas
are summarized in Table \ref{table:ant-perf}.  Despite the different
antenna sizes and primary illumination radii, the main beams and first
sidelobe levels are identical for both antennas. The cross polar
signal levels are also very similar. The effect of the primary and
secondary baffles is shown in simulation in Fig. \ref{fig:baffles}. Here the results are shown at 4.5 GHz, since
the effect is strongest at the lower edge of the observing band. The
sidelobes at 50$^\circ$ are due to spillover past the secondary mirror
and are suppressed by the baffle around the secondary. The spillover
past the primary at about 110$^\circ$ is suppressed by the primary
baffle.

The secondary baffle is implemented by wrapping foil-backed absorber
(Advanced ElectroMagnetics AEL-0.375) around the cylindrical section
of the foam secondary support structure. For the primary baffle, a
structure of curved aluminium sheets supported on struts connected to
the outer rim of the primary reflector was constructed, and the
absorber glued to the inner face of the sheets.

\section{Antenna performance}

%
\begin{table}[!t]
\renewcommand{\arraystretch}{1.3}
\caption{Antenna Characteristics and Simulated Performance}
\label{table:ant-perf}
\centering
\begin{tabular}{|>{\centering}m{4cm}|c|c|}
\hline
Performance at 5~GHz & 6.1-m antenna & 7.6-m antenna \\
& & \\
\hline
FWHM & 0.73$^\circ$ & 0.73$^\circ$\\
\hline
Gain & 47.6 dBi  & 47.7 dBi \\
\hline
First sidelobe level & $-$19.1 dB & $-$19.3 dB \\
\hline
Cross-polar level & $-$51 dB & $-$52 dB \\
\hline
Primary mirror radius  & 3.048~m & 3.835~m \\
\hline
Primary illumination radius ($-40$~dB) & 2.96~m & 3.25~m \\
\hline
Secondary mirror radius & 0.51~m & 0.50~m \\
\hline
Antenna type & Gregorian & Cassegrain \\
\hline
Primary baffle depth & 0.80~m & -- \\
\hline
Secondary baffle depth & 0.30~m & -- \\
\hline
Primary baffle temp. contribution & 0.7~K & -- \\
\hline
Secondary baffle temp. contribution & 0.2~K & -- \\
\hline
Beam efficiency (power within $\pm $5$^{\circ}$) & 91.9\% & 91.3\% \\
\hline
Main beam efficiency (power within first null)& 80.0\% & 80.0\% \\
\hline
\end{tabular}
\end{table}

\subsection{Main beam measurements}

\begin{figure}[!t]
\center
\includegraphics[width=3.5in]{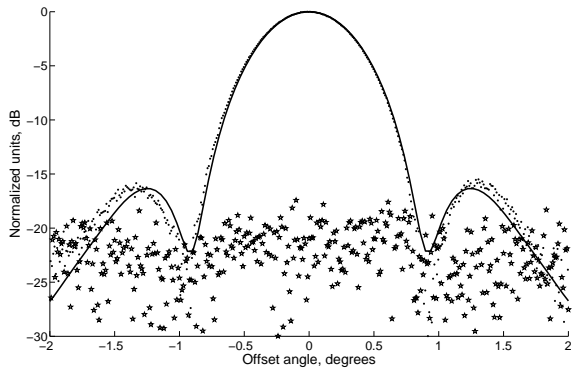}
\caption{Simulated main beam (solid line), averaged over the system
  bandwidth, and measured beam (dots) measured from azimuth scans on
  the astronomical source Cas A. Stars show the
    polarized ($P = (Q^2 + U^2)^{1/2}$) response to this unpolarized
    source, after subtracting a leakage term due to imperfections in the
    polarimeter. This is an upper limit on the degree of
    cross-polarization in the optics -- the expected level is $< -50
    \, \rm dB$.}
\label{fig:beam-map}
\end{figure}

The 6.1-m antenna can only be pointed to a minimum
  elevation of $13^{\circ}$, and hence cannot be easily pointed to an
  artificial source in the far field. We therefore measured the main
  beam using an astronomical source and the radiometer backend built
for the C-BASS survey, which measures the difference between the power
received by the feed horn and that from a stabilized cold load. The
antenna was repeatedly swept backwards and forwards in azimuth about
the position of the radio source Cassiopeia A (Cas A), which is one of
the brightest compact radio sources in the northern sky (670 Jy at 5
GHz \cite{Hafez_2008}). Cas A is approximately 5 arcmin in diameter,
and thus much smaller than the beam (FWHM 44 arcmin). A linear fit was
removed from each scan to remove the effect of changing airmass during
the observation and also to remove the diffuse Galactic emission near
Cas A (around 1 per cent of the peak source brightness). The data were
then stacked as a function of offset from the nominal source
position. Fig. \ref{fig:beam-map} shows the stacked data, overlaid
with the simulated main beam lobe and first sidelobes. 

At the time of
these observations, the actual position of the secondary mirror was
measured to be raised 5 mm higher than nominal, so for direct
comparison the simulation of the main beam was recalculated with this
offset included in the geometry. The simulated beam was calculated at
three frequencies, 4.5, 5.0 and 5.5 GHz. We modeled the variation of
the flux density of Cas A with frequency as $\nu^{\alpha}$, where
$\nu$ is the frequency and $\alpha$ is the spectral index, which for
Cas A is $-$0.767 \cite{Hafez_2008}. The simulated beam was formed by
weighting each narrow-band simulation by the expected flux density and
then averaging over the three frequencies. Fig. \ref{fig:beam-map}
shows excellent agreement between the simulated beam and the
measurement of the main beam and first sidelobe.  

We also used these measurements to place a limit on the cross-polar
  levels of the main beam.  In Fig \ref{fig:beam-map} we show the
  measured polarization ($P = (Q^2 + U^2)^{1/2}$ ) response to the
  (nominally unpolarized) source Cas~A. We subtracted the best-fitting
  scaled co-polar beam from the scanned $P$ data, which represents
  polarization leakage ($I \rightarrow P$) in the polarimeter, plus
  any intrinsic polarization of the source. The appropriate scaling
  was approximately 1\% of the co-polar peak.  Cross-polarization in
  the optics is expected to have a quite different beamshape, with a
  null at the centre and peaks near the first null of the co-polar
  beam. No such beamshape is evident in the plots at the level of
  $\sim -20 \, \rm dB$ down from the peak. The expected level of
  cross-polarization is $< -50\, \rm dB$ relative to the peak co-polar
  response (see Fig \ref{fig:near-beams}), much smaller than this
  measured limit. With the southern 7.6-m telescope, which can be
  pointed to the horizon and hence to an artificial source, we expect
  to make a much more sensitive measurement of the optical
  cross-polarization.

\subsection{Far-out sidelobe measurements}

\begin{figure}[!t]
\center
\includegraphics[width=3.5in]{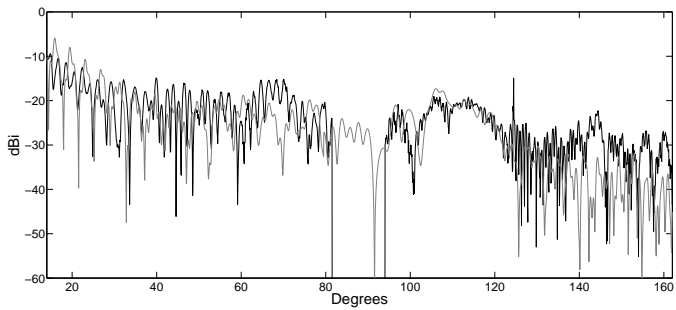}
\caption{Measured (black) and simulated (grey) beam pattern of the 6.1m
  antenna without baffles between $13^{\circ}$ and $163^{\circ}$ at
  5~GHz. There is no measured data between $82^{\circ}$ and
  $95^{\circ}$. Note that the vertical axis is in dBi and the peak of
  the main beam is at $+47.6$~dBi.}
\label{fig:far-comparison}
\end{figure}

For far-out sidelobe measurements, a ground-based 5 GHz CW signal
source was used at a distance of 4.5 km, in the antenna's far
field. Due to pointing limitations of the telescope mount we were only
able to map the sidelobe pattern between boresight angles of
13$^\circ$ -- 82$^\circ$ and 95$^\circ$ -- 163$^\circ$. The signal source was switched on and off
at 0.7 Hz to allow discrimination between the test signal and ground
pick-up and background sources. Fig. \ref{fig:far-comparison} shows
the beam pattern measurement compared to the simulation results at 5
GHz, without baffles. Since the exact distance and power
level of the signal source was known, the absolute gain of the antenna
pattern in dBi could be calculated. Fig. \ref{fig:far-comparison}
shows very good agreement between measurement and simulation even at
signal levels of more than 60 dB below the antenna's forward gain.

\begin{figure}[!t]
\center
\includegraphics[width=3.5in]{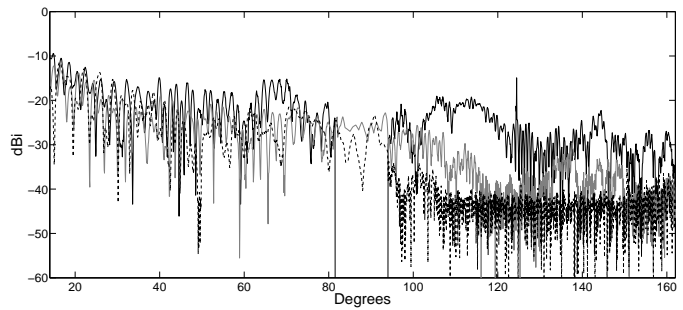}
\caption{Three far-out sidelobe measurements between $13^{\circ}$ and $163^{\circ}$ at 5 GHz without baffle (black), with secondary baffle
  only (grey) and with both baffles (dotted black).}
\label{fig:far-baffles}

\end{figure}

Two further measurements were conducted including the absorbing
baffle around the secondary, and around both the secondary and
primary, as discussed in section III. Fig. \ref{fig:far-baffles}
compares the three measurements with and without the absorbing
baffles. There is a clear reduction of sidelobe levels at 50$^\circ$
and 110$^\circ$. The backlobe response (at angles greater than
100$^\circ$) is reduced by 10--20 dB and the absolute level is reduced
to more than 90 dB below the main beam. Measurements were made both
with the original aluminium secondary mirror and LD foam cone, and
with the replacement carbon-fibre mirror and HD foam cone, with
identical results.

\section{Conclusions}

We have presented the development of two circularly-symmetric radio
antennas which have been modified to meet the special needs of high
polarization purity and low signal contamination for an all-sky survey
at 5~GHz. These needs have been met by three key design
features. First, a low sidelobe and low cross-polarization feed horn
has been developed.  Second, the secondary mirror is supported by a
transparent circular Zotefoam cone without metal struts, and third,
absorbing baffles have been fitted to minimize spillover. We have
shown simulated results and compared these to measurements of one of
the two antennas, the 6.1-m antenna at the Owens Valley Radio
Observatory. In particular, the foam secondary support cone makes only a small contribution to the system temperature and is a good option to conserve
polarization purity in a circular symmetric design without the need
for metal support struts. The use of absorber around the rims of both
the primary and secondary mirrors is shown to decrease significantly
the sidelobes due to spillover around both mirrors.

\section*{Acknowledgment}

The authors would like to thank the staff of the Mechanical
Engineering group in the Oxford University Department of Physics, particularly Mike Tacon, Paul Rossiter and Matthew Brock,  for
the construction of the C-BASS optical components. We also thank the staff of
the Owens Valley Radio Astronomy Observatory, particularly Russ
Keeney, for support of the operations of the C-BASS North telescope.

\ifCLASSOPTIONcaptionsoff
  \newpage
\fi



%
\bibliographystyle{IEEEtran}
\bibliography{IEEEabrv,cbass}

\begin{thebibliography}{10}
\providecommand{\url}[1]{#1}
\csname url@samestyle\endcsname
\providecommand{\newblock}{\relax}
\providecommand{\bibinfo}[2]{#2}
\providecommand{\BIBentrySTDinterwordspacing}{\spaceskip=0pt\relax}
\providecommand{\BIBentryALTinterwordstretchfactor}{4}
\providecommand{\BIBentryALTinterwordspacing}{\spaceskip=\fontdimen2\font plus
\BIBentryALTinterwordstretchfactor\fontdimen3\font minus
  \fontdimen4\font\relax}
\providecommand{\BIBforeignlanguage}[2]{{%
\expandafter\ifx\csname l@#1\endcsname\relax
\typeout{** WARNING: IEEEtran.bst: No hyphenation pattern has been}%
\typeout{** loaded for the language `#1'. Using the pattern for}%
\typeout{** the default language instead.}%
\else
\language=\csname l@#1\endcsname
\fi
#2}}
\providecommand{\BIBdecl}{\relax}
\BIBdecl

\bibitem{cbass_king}
O.~G. King \emph{et~al.}, ``{The C-Band All-Sky Survey: instrument design,
  status, and first-look data},'' in \emph{Society of Photo-Optical
  Instrumentation Engineers (SPIE) Conference Series}, vol. 7741, 2010.

\bibitem{WMAP}
C.~L. Bennett \emph{et~al.}, ``{The Microwave Anisotropy Probe Mission},''
  \emph{ApJ}, vol. 583, pp. 1--23, Jan. 2003.

\bibitem{Planck}
{Planck Collaboration}, ``{Planck Early Results: The Planck mission},''
  \emph{A\&A}, vol. 536, p.~A1, 2011.

\bibitem{Imbriale}
W.~A. Imbriale and R.~Abraham, ``Radio frequency optics design of the deep
  space network large array 6-meter breadboard antenna,'' Jet Propulsion
  Laboratory, Pasadena, CA, The Interplanetary Network Progress Report, vol.
  42-157, May 15 2004.

\bibitem{Galindo_1964}
V.~Galindo, ``Design of dual-reflector antennas with arbitrary phase and
  amplitude distributions,'' \emph{IEEE Trans. Antennas and Propagation},
  vol.~12, no.~4, pp. 403--408, 1964.

\bibitem{reich82}
W.~{Reich}, ``{A radio continuum survey of the northern sky at 1420 MHz. I},''
  \emph{A\&AS}, vol.~48, pp. 219--297, May 1982.

\bibitem{Cruckshank_2007}
P.~S. Cruckshank, D.~R. Bolton, D.~A. Robertson, R.~J. Wylde, and G.~M. Smith,
  ``Removing standing waves in quasi-optical systems by optimal feedhorn
  design,'' in \emph{IRMMW-THZ Conference Digest}, Sep 2007, pp. 941--942.

\bibitem{corrug}
\BIBentryALTinterwordspacing
Corrug. SMT Consultancies Ltd. [Online]. Available:
  \url{http://www.smtconsultancies.co.uk/products/corrug/corrug.php}
\BIBentrySTDinterwordspacing

\bibitem{Keating}
\BIBentryALTinterwordspacing
{Thomas Keating Ltd}, ``{Ultra-Gaussian Corrugated Horns},'' Tech. Rep.
  [Online]. Available:
  \url{http://www.terahertz.co.uk/index.php?option=com_content&view=article&id=189&Itemid=567}
\BIBentrySTDinterwordspacing

\bibitem{Grimes_2007}
P.~K. Grimes, O.~G. King, G.~Yassin, and M.~E. Jones, ``Compact broadband
  planar orthomode transducer,'' \emph{Electronic Letters}, vol.~43, pp.
  1146--1148, 2007.

\bibitem{CBI2_2011}
A.~C. Taylor \emph{et~al.}, ``{The Cosmic Background Imager} 2,'' \emph{MNRAS},
  vol. 418, pp. 2720--2729, 2011.

\bibitem{Dybdal_1979}
R.~Dybdal and H.~King, ``Performance of reflector antennas with absorber-lined
  tunnels,'' in \emph{Antennas and Propagation Society International Symposium,
  1979}, vol.~17, Jun 1979, pp. 714 -- 717.

\bibitem{Hafez_2008}
Y.~A. Hafez \emph{et~al.}, ``Radio source calibration for the very small array
  and other cosmic microwave background instruments at around 30 {GHz},''
  \emph{MNRAS}, vol. 388, pp. 1775--1786, 2008.

\end{thebibliography}



\begin{IEEEbiographynophoto}{Christian M. Holler}
was born in southern Germany in 1973. He studied physics at the Ludwig-Maxmilians-Universit\"{a}t in Munich, Germany and received his PhD in the field of astrophysics and instrumentation at the Cavendish Laboratories, Cambridge University, UK in 2003.  After successfull entrepreneurial work he joined the experimental cosmology group at the University of Oxford as a postdoctoral research assistant, where he was involved in several instrumentation projects. In September 2009 he was appointed a Professor at the University of Applied Sciences in Esslingen, Germany.
\end{IEEEbiographynophoto}

\begin{IEEEbiographynophoto}{Angela C. Taylor} was born in Dartford, UK in 1975 and studied Natural Sciences at Cambridge University, earning a PhD in astrophysics at the Cavendish Laboratory Cambridge in 2002. She continued to work in the field of astrophysics and instrumentation as a postdoctoral research assistant at the Cavendish Laboratory until 2005 when she moved to the University of Oxford as a founding member of the experimental cosmology group. Since January 2005 she has been an STFC Postdoctoral Research Fellow and then a Royal Society Dorothy Hodgkin Research Fellow working on a range of experimental cosmology projects at the University of Oxford.
\end{IEEEbiographynophoto}

\begin{IEEEbiographynophoto}{Michael E. Jones} 
was born in Wrexham, UK, in 1966 and studied Natural Sciences and Electrical Science at Cambridge University. He obtained a PhD in Radio Astronomy at the Cavendish Laboratory, Cambridge in 1990 and subsequently worked there as a Research Associate and Senior Research Associate. In 2005 he was appointed first as a Lecturer and then as Professor of Experimental Cosmology at the University of Oxford Department of Physics. 
\end{IEEEbiographynophoto}

\begin{IEEEbiographynophoto}{Oliver G. King} was born in Bedford, South Africa in 1983. He studied Physics and Electronics at Rhodes University in Grahamstown, South Africa. He obtained a DPhil in Astrophysics at the University of Oxford in 2009, where he worked on radio instrumentation. He moved to the California Institute of Technology in Pasadena, California in late 2009 to take up a Keck Postdoctoral Fellowship. At Caltech he leads astrophysics instrumentation projects from radio to optical wavelengths.
\end{IEEEbiographynophoto}

\begin{IEEEbiographynophoto}{Stephen J. C. Muchovej} was born in southeastern Brasil in 1979.  He earned a Bachelors degree in physics and astronomy from the University of California, Berkeley.  He earned his PhD in 2008 from Columbia University in the field of astrophysics with a focus on instrumentation and experimental cosmology.  He is currently a National Science Foundation fellow at the California Institute of Technology. 
\end{IEEEbiographynophoto}

\begin{IEEEbiographynophoto}{Matthew A. Stevenson} was born in Winnipeg, Manitoba, Canada, in the early 1980s. He received his BSc in Astronomy and Physics from University of Victoria, British Columbia, Canada. Since 2006 he is a PhD student in Astronomy at the California Institute of Technology.
\end{IEEEbiographynophoto}

\begin{IEEEbiographynophoto}{Richard J. Wylde} was born in London in 1958 and educated at Eton and Sidney Sussex College, Cambridge and studied Natural Sciences. He joined Queen Mary College, University of London, and received the Ph.D. degree in physics in 1985. He is a Fellow of the Institution of Engineering and Technology. He is currently an Honorary Reader in the School of Physics and Astronomy, University of St. Andrews, Scotland, and he manages both Thomas Keating Ltd. precision engineers, and QMC Instruments Ltd. 
His research interests lie in the design of THz systems for use in thermo-nuclear plasma diagnostics, observational cosmology, astronomy, electron-spin-resonance spectroscopy and atmospheric remote sensing.
\end{IEEEbiographynophoto}

\begin{IEEEbiographynophoto}{Charles Copley} was born in Pietermaritzburg, South Africa in 1982 and educated at Michaelhouse, Rhodes University and the University of Cape Town. Since 2010 he has been pursuing a DPhil at the University of Oxford, with a focus on instrumentation.
\end{IEEEbiographynophoto}

\begin{IEEEbiographynophoto}{Richard J. Davis} was born in March, UK in 1949 and studied Natural Sciences at Cambridge University, earning a PhD in radio astronomy  at the
University of Manchester in 1975. He continued to work in the field
of astrophysics and instrumentation as an SERC fellow and then a University
PDRA  at Jodrell Bank till 1978 and then as an academic member of staff.
\end{IEEEbiographynophoto}

\begin{IEEEbiographynophoto}{Timothy J. Pearson} is a senior research associate in radio astronomy at the California Institute of Technology. He received his Ph.D. from the University of Cambridge in 1977.  His research interests include radio astronomy, interferometry, active galactic nuclei, and the cosmic microwave background radiation.
\end{IEEEbiographynophoto}

\begin{IEEEbiographynophoto}{Anthony C. S. Readhead} is the Barbara and Stanley R. Rawn, Jr., Professor of Astronomy at the California Institute of Technology, and Director of the Owens Valley Radio Observatory and a Senior Research Scientist at the Jet Propulsion Laboratory. He studied physics and mathematics at the University of the Witwatersrand in South Africa, and conducted Ph.D. research on active galaxies and interplanetary scintillation at the Cavendish Laboratory, University of Cambridge. He was elected to the National Academy of Sciences and the American Academy of Arts and Sciences in 1995.
\end{IEEEbiographynophoto}

\end{document}